\documentclass[a4paper]{jpconf} 
\usepackage{graphicx}
\usepackage[ruled,vlined]{algorithm2e}
\begin{document}
\title{A study on Channel Popularity in Twitch}

\author{Ha Le, Junming Wu, Louis Yu and Melissa Lynn}

\address{Gustavus Adolphus College}

\ead{\{hle5, jwu, lyu, mklynn\}@gustavus.edu}

\begin{abstract}
In the past few decades, there has been an increasing need for Internet users to host real time events online and to share their experiences with live, interactive audiences. Online streaming services like Twitch have attracted millions of users to stream and to spectate. There have been few studies about the prediction of streamers' popularity on Twitch. In this paper, we look at potential factors that can contribute to the popularity of streamers. Streamer data was collected through consistent tracking using Twitch's API during a 4 weeks period. Each user's streaming information such as the number of current viewers and followers, the genre of the stream etc., were collected. From the results, we found that the frequency of streaming sessions, the types of content and the length of the streams are major factors in determining how much viewers and subscribers streamers can gain during sessions.
\end{abstract}

\section{Introduction}
Social Media has made enormous impacts on how we share information and experiences with each other. There exists many social networking platforms that allow us to post images, videos, and other content. In the past few years, as the video games and e-sports industry grows rapidly, generating billions dollar of revenue from viewership \cite{borowy2013pioneering}, it has become clear that there is a need for players to share gaming experiences with one another. Since then, many gaming platforms have gained momentum in streamers and viewership. Among them Twitch has stood out as the best-known for streaming videos games and broadcasting major e-sports events.

Twitch is the most popular live streaming platform worldwide, which draws more than 100 millions of viewers every month. Even though Twitch attracted streamers from multiple categories, the main portion of its audience and streamers are competitive players and video game streamers. It stands apart from other mainstream platforms by providing streamers various opportunities to build a strongly connected online communities. Streamers on Twitch has shown to have enormous impact on the gaming industry \cite{johnson2019impacts}. Likewise, the donation from audience and partnership with Twitch can financially support gamers to become full-time, professional video game streamers. The race to attract viewers has become increasingly competitive on the platform, especially for beginner streamers. Thus, finding key elements to a fast-growing, successful Twitch channel is crucial. Even though some research has attempted to analyze Twitch communities \cite{10.1145/2556288.2557048,Churchill}, few has examined the quantitative factors that contribute to the popularity of Twitch channels and their corresponding streamers.

In this paper, we collect data by crawling to explore the distribution of popularity among streams. We categorize them and conduct analysis in order to identify characteristics of successful streamers. We have concluded that a successful streaming strategy includes: (1) streams do not exceed 5 hours, (2) at least 5 streams per week and (3) combine non-gaming and gaming content. Applying those characteristics, we also predict the growth rates of channels through machine learning classification model.

The rest of this paper is organized as follows. In Section 2, we provide some background information. In Section 3, we survey related work. Section 4 describes the setup of our experiments, followed by results in analysis in Section 5. Finally, we give our conclusion in Section 6.

\section{Background}
\subsection{Social Media}
Online social media has grown tremendously over the past few years. According to a study in 2016 \cite{greenwood2016social}, 68$\%$ of the adults in the United States use the Internet and 79$\%$ of them engage in at least one form of social networking platforms. Facebook, a major online social networking platform, announced that they have more than one billion users around the world in 2012, while LinedkIn reported to have half a billion users in 2021 \cite{dewing2010social}. Social media platforms have created online communities that allow people to make and share content, to make connections and to communicate with each other. There has been research on online communities's effects on our political \cite{qin2016political}, social \cite{o2011impact,siddiqui2016social} and economical \cite{edosomwan2011history} believes.

\subsection{Streaming Services}
Streaming media is a booming industry. The main goal of these services is to deliver continuous and uninterrupted content to their users. Streaming platforms usually have two components: the component the broadcasts content and s comments/audience interaction Section. The streaming content on these platforms is diverse. Today major social media platforms such as Facebook, Instagram, YouTube, etc. have adopted streaming components. The most popular streaming services are for broadcasting movies and TV shows. This is a very competitive market with new platforms developed every year. For example, Netflix, HBO Max, Disney+, Amazon Prime Video, YouTube etc. Music streaming is also a large component of the industry, in which Spotify is the leading platform. Many people uses the streaming components in social networking platforms such as Facebook and Instagram for sharing experiences, interacting with followers, and sometime selling products and advertising. 
\subsection{Twitch}
Established in 2011 and acquired by Amazon in 2014 for \$974,  Twitch focus primarily on video game and e-sports has become the leading streaming service provider\cite{BBC}\cite{livereport}. In 2020, Twitch has attracted over 30 million average viewers daily, and over 4.5 billion hours of content were watched on Twitch during the third quarter. Differ from other platform, primary users on Twitch are young adolescent and adult men who is also the dominant gamer\cite{gamerdemographic}. Therefore, Twitch provides an ideal environment for streamers reaching video game related audiences.

The live streaming environment on Twitch facilitates direct engagement between the streamer and the viewers. Streamers and audiences can interact through a live chat box nearby the broadcast, while other social media such as Instagram or YouTube only provide a time-lagged comments sections for audiences to reach out creators. Twitch allows viewers to directly support their favorite streamers by buying and giving tokens. For avid viewers, they can choose to subscribe to their favorite streamers and receive corresponding perks provided by the channels. In order to attract more fans and supports, streamers need a more sophisticated communication techniques during live-streaming than simply posting pictures or videos on other social media. Due to the live-streaming environment, Twitch users also possess strong feeling of connectivity with other viewers \cite{Tim2020}. Research has shown that people are able to develop meaningful, maintainable relationships through the Twitch platform \cite{10.1145/3415165}. Such connection may also be associated with a history of monetary donation to Twitch content creators \cite{Zorah2018}.

\section{Related Works}
\subsection{Studies on Streaming Services: }

Pires and Simon \cite{pires2015youtube} presented a 3-months dataset on Youtube and Twitch. Their preliminary studies show that both platforms, especially Twitch, are able to generate rich services at any time. The popularity of the channel is more heterogenous than what have been observed in other social media platforms. Zhang et al. \cite{zhang2013understanding} conducted a survey on consumer behaviors on Spotify, a popular music streaming platform. They identified some key characteristics on user behaviors; this includes the users' favorite time of the day, the length of a successive user session and the tendencies to switch devices. Lee et al. \cite{lee2016cannibalizing} examined the impact of music streaming services on record sales in South Korea and showed that music streaming has a positive impact on the record sales. Surprisingly, the number of songs in an album and the number of singers in a group also have an impact on record sales as well.

\subsection{Studies on Streaming Communities: }

Gandhi et al. \cite{gandhi2021exploration} discussed how streamers make connections with their audiences. Todd and Melacon \cite{todd2018gender} explored the impact of streamers' genders on the perception of viewers in live streaming sessions and found that the likelihood of a viewer interacting with streamers depend heavily on the sex of the streamers. Cai et al. \cite{cai2021understanding} looked at micro-communities in streaming platforms and showed that there are correlations between the imposed regulations, the vibe of the channels, and harassment tendencies within the community. Cai and Wohn \cite{cai2021moderation} examined the role of volunteer moderators in streaming communities. 

\subsection{Studies on Twitch: }
Although we cannot find research similar to ours, there have been other studies on Twitch. Seering et al. \cite{seering2018social} looked at chatbots, programs designed to perform automated tasks during streams, and showed how they can help communities grow and evolve. Poyane \cite{poyane2019toxic} looked at chat logs from Twitch channels to create a topic model of viewers discussions.

\subsection{Twitch Users' Profiles: }
Many researchers have analyzed the profile of Twitch users. Kim et al. \cite{kim2019subscriber} looked at users' differences in sentiments through messages in chat; from that, they attempted to identify potential subscribers' characteristics. Lee \cite{lee2019study} approached the same topic of whether a viewer will subscribe to a channel through machine learning, building models such as logistic regression, SVM, decision tree and random forest. Han and Sukhee \cite{han2016characteristics} collected data to examine the distinctions between channels in South Korea. Zhao et al. \cite{Zhao} attempted to identify human factors such as openness, personal affordance and gaming levels that can contribute to the popularities of channels. 

\subsection{Predicting the Popularity of Users on YouTube, Twitter, and Spotify: }
While there are few research on the popularity of users on Twitch, there are studies on other streaming platforms such as YouTube, Twitter, and Spotify. Figueiredo, Benevenuto and Almeida \cite{FigueiredoYouTubePatterns} studied the growth patterns of video popularity on YouTube, Based on the data collected, they categorized the videos into three categories and observed the characteristics of videos that tend to motivate viewers into watching, thus contribute to channel growths. Pinto, Almeida, and Goncalves \cite{PintoYouTubePrediction} present two simple models for predicting the popularity of Web content on YouTube based on historical information given by popularity measures. Ma, Yan and Chen \cite{MaYoutubeLongterm} achieved fast prediction of long-term video popularity on YouTube. To understand the propagation of news Tweets, Wu and Shen \cite{wu2015analyzing} built a model that can predict the final number of retweets for a news tweet based on information collected from super nodes on Twitter. Gao et al. \cite{gao2019taxonomy} studied popularity prediction of posts in microblogs. Hu and Sci \cite{yu2020prediction} looked at predicting peak time popularity of tweets based on Twitter hashtags. 

Studies have been done on the prediction of user popularities on Spotify, one of the most popular music streaming platforms. Araujo, Cristo and Giusti \cite{Soares2019Classify} approached this problem as a classification task and employed classifiers built on past information from the platform's Top 50 Global ranking as well as acoustic features from the songs. They found that RBF kernel gave the best results with an AUC higher than 80$\%$ when predicting the popularity of songs two months in advance. Wu and Sun \cite{9434816} adopted Principle Components Analysis (PCA) and model blend-in, avoiding significant error by high dimension,  and gave a relatively precise prediction on the popularity of Spotify's top music. 

\section{Experiment Setup}
Our data set consists of streaming information from 20,000 random active Twitch users, gathered using Twitch's API. Additional information on the games themselves was crawled from the web using a web crawler. The Twitch API does not provide achieved streaming information so all streams were real-time data obtained by consistently tracking activities from users over the course of two month. We start by generating a dataset of 10,000 active Twitch users by requesting 20,000 random streams currently on the platform and extracting the ids from the streamers. 
\subsection{Gathering Streaming Information:}

Over the course of 4 weeks, we closely monitor the activities of the selected accounts. Every week we choose 4,000 users and monitor their activities; every 30 minutes we record their current status and their number of followers. If the account is streaming, we record the id of the stream as well as the current number of viewers, the game tag and the streaming language. After one week, we switch to another set of 4,000 users and repeat the process. 

\subsection{Gathering Additional Information:}

Apart from the real-time information gathered from live streams on Twitch, we also looked for additional information related to the channel. Specifically, we gathered information on any additional videos and clips uploaded to the channels after the streams have finished. That is, after streams have ended, the users can upload videos cut from parts of the streams to their channel. Using the streams' ids, we were able to track them. If there were special moments during the stream that have the potential to go viral after, streamers can choose to cut a short clip from the stream and upload it as highlight. Using the streams' ids and the videos' ids, we were able to track the highlights. 

We think that the popularity of games played during streams is also an important factor on the views. Using a web crawler, we obtained daily popularity of each game title using Twitch's self-reported statistics. 

Over the course of one month, we were able to get the complete data of 10,058 users, with more than 55000 streams and 50000 video clips. Our dataset consists of streams in 20 different languages, with both gaming and non-gaming contents. 

\section{Analysis}

\subsection{The Power Law Distributions}

First, we look at the popularity of streamers in the dataset by examining the distributions of their corresponding number of followers and viewers. Figure 1, 2, 3 shows the distribution of followers, the distribution of average viewers during streams, and the distribution of viewers tuning in at the start of the stream. All three follow the power law with exponents 1.25, 1.11 and 2. This indicates an underlying relationship between the popularity of streamers and their ability to attract followers/viewers. That is, streamers' popularity will grow at a rate proportional to his/her current followers/viewers count.
\begin{figure}[h]
\begin{minipage}{16pc}
\includegraphics[width=14pc]{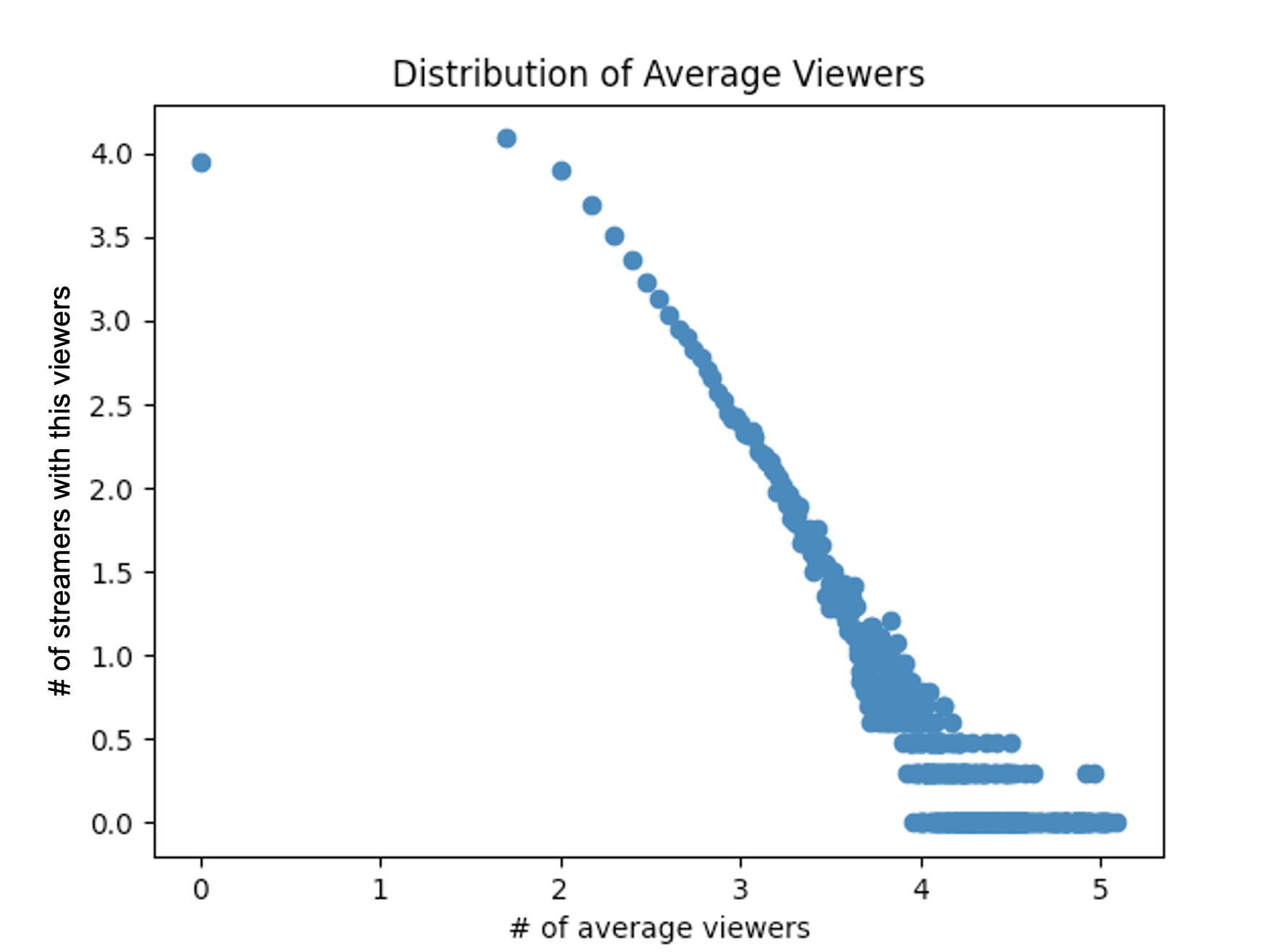}
\caption{Distribution of Average viewers}
\end{minipage}\hspace{2pc}
\begin{minipage}{16pc}
\includegraphics[width=14pc]{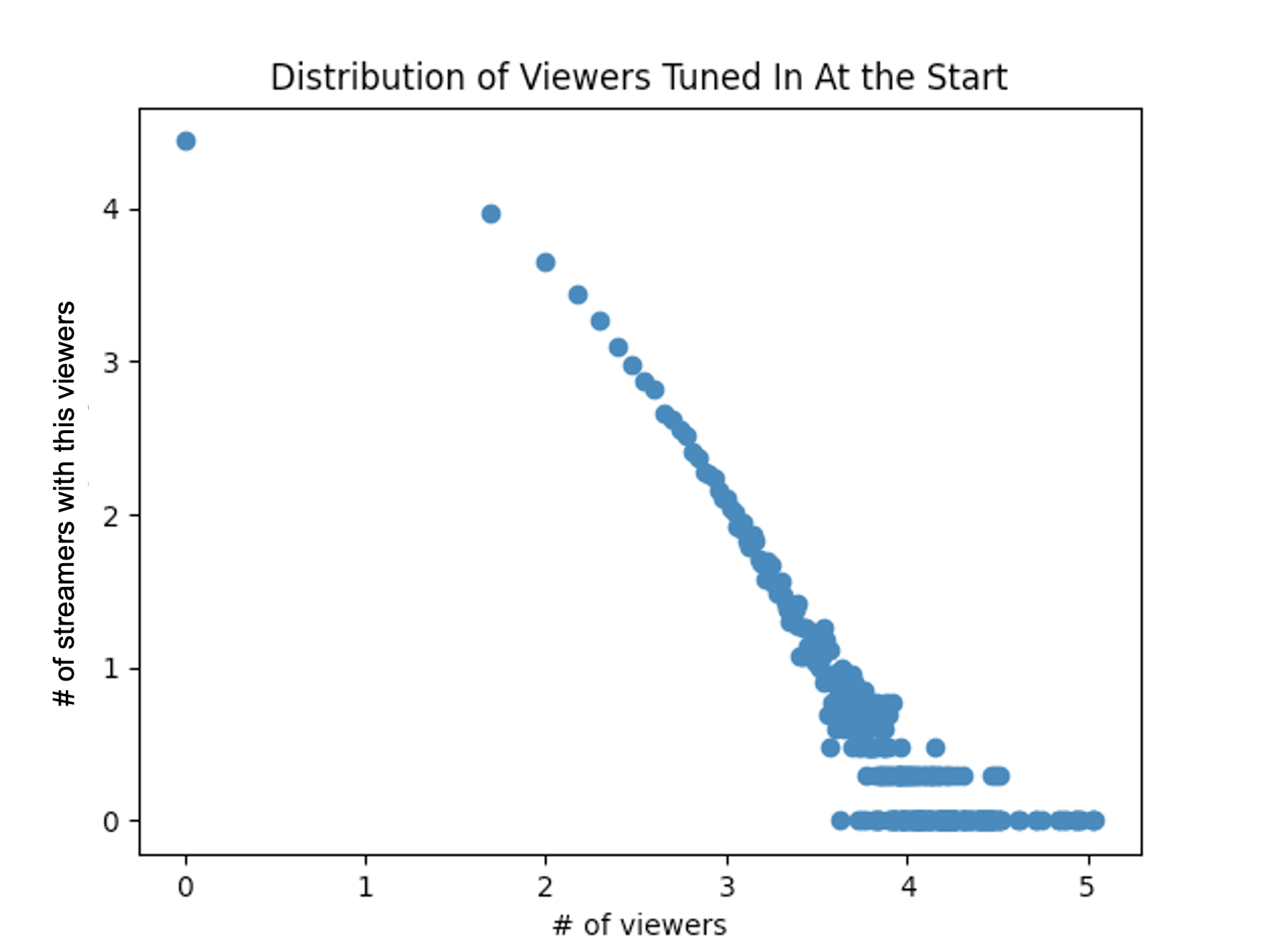}
\caption{Distribution of Viewers at the Start}
\end{minipage} \hspace{2pc}
\begin{minipage}{16pc}
\includegraphics[width=14pc]{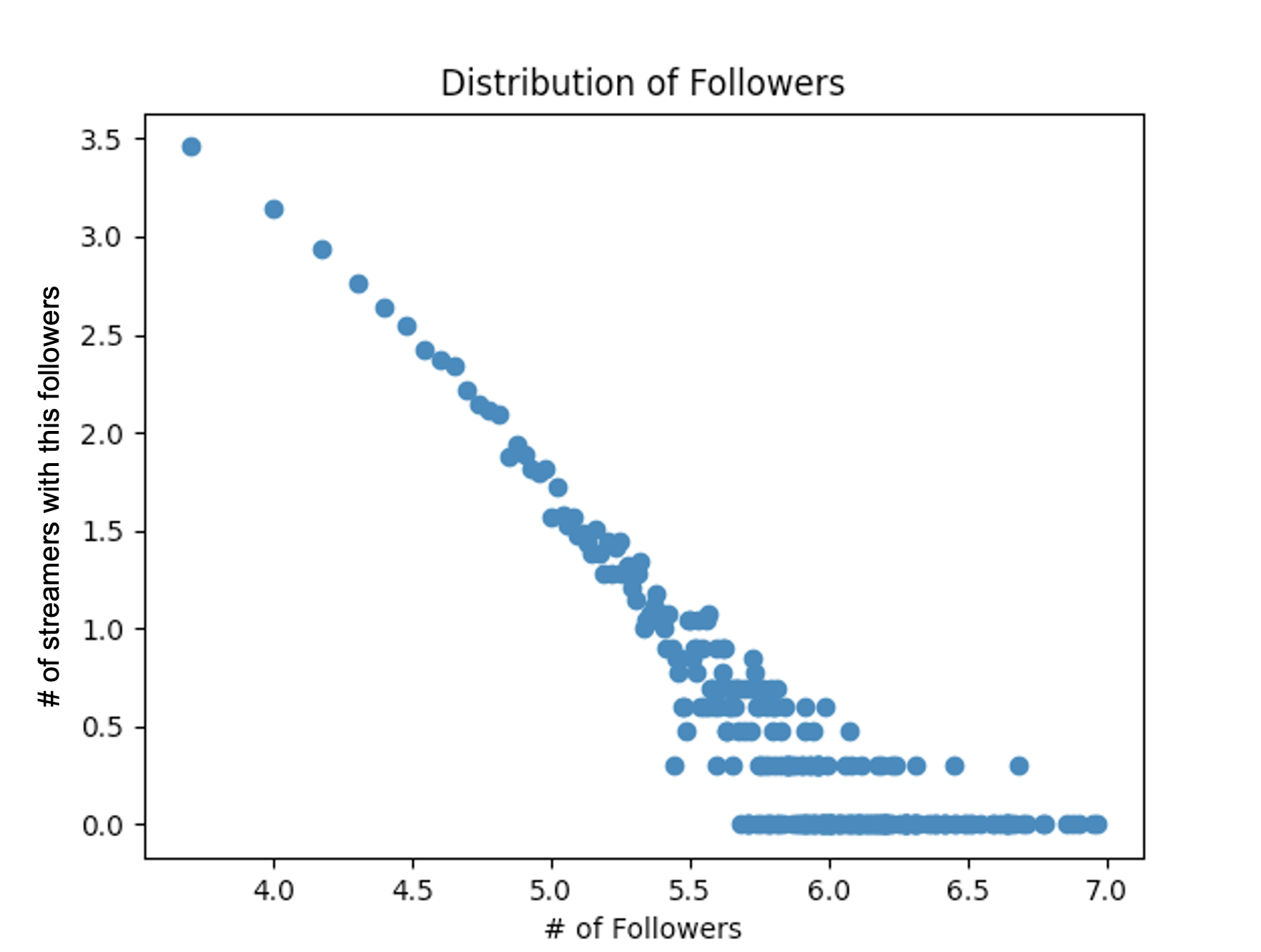}
\caption{Distribution of Followers}
\end{minipage} 
\end{figure}

\subsection{Distribution of Followers Gained and Bot Detection}
Next, we looked at the average distribution of followers gained during the week. Users' activities during the week are categorized as active and inactive. Figure 4, 5 shows the distribution of followers gained during streaming (active) period, and the distribution of followers gained during non-streaming (inactive) period. We observed that the graphs are spiky. They also suggest that at times streamers were getting more followers during inactive periods than during active periods. These results are curious to us. A closer look at the data reveals several suspicious surges and declines in the number of followers for many streamers. We suspect bot interference from these abnormal behaviors. 

\begin{figure}[h]
\begin{minipage}{16pc}
\includegraphics[width=14pc]{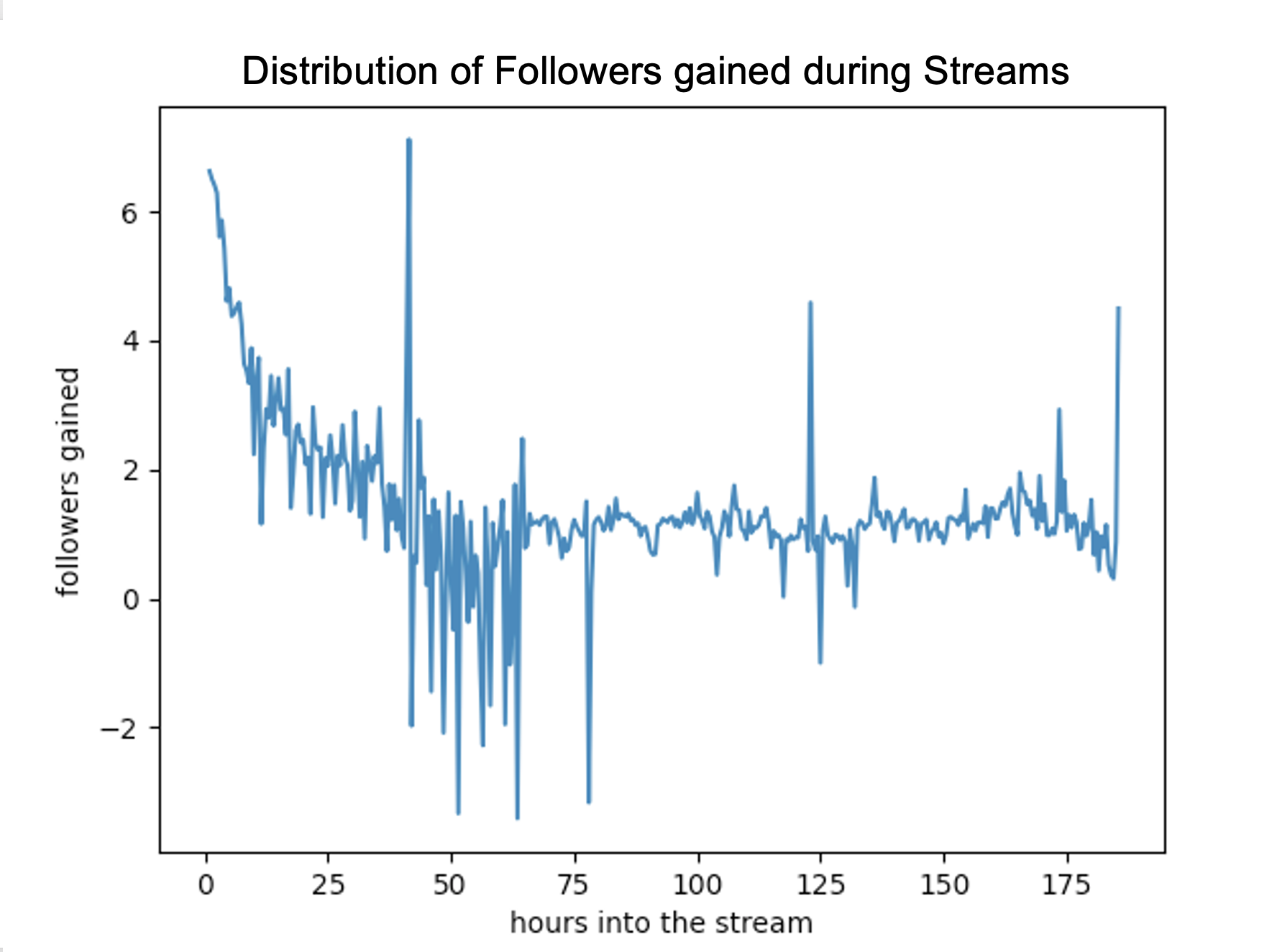}
\caption{Distribution of Followers Gained Active Period}
\end{minipage}\hspace{2pc}
\begin{minipage}{16pc}
\includegraphics[width=14pc]{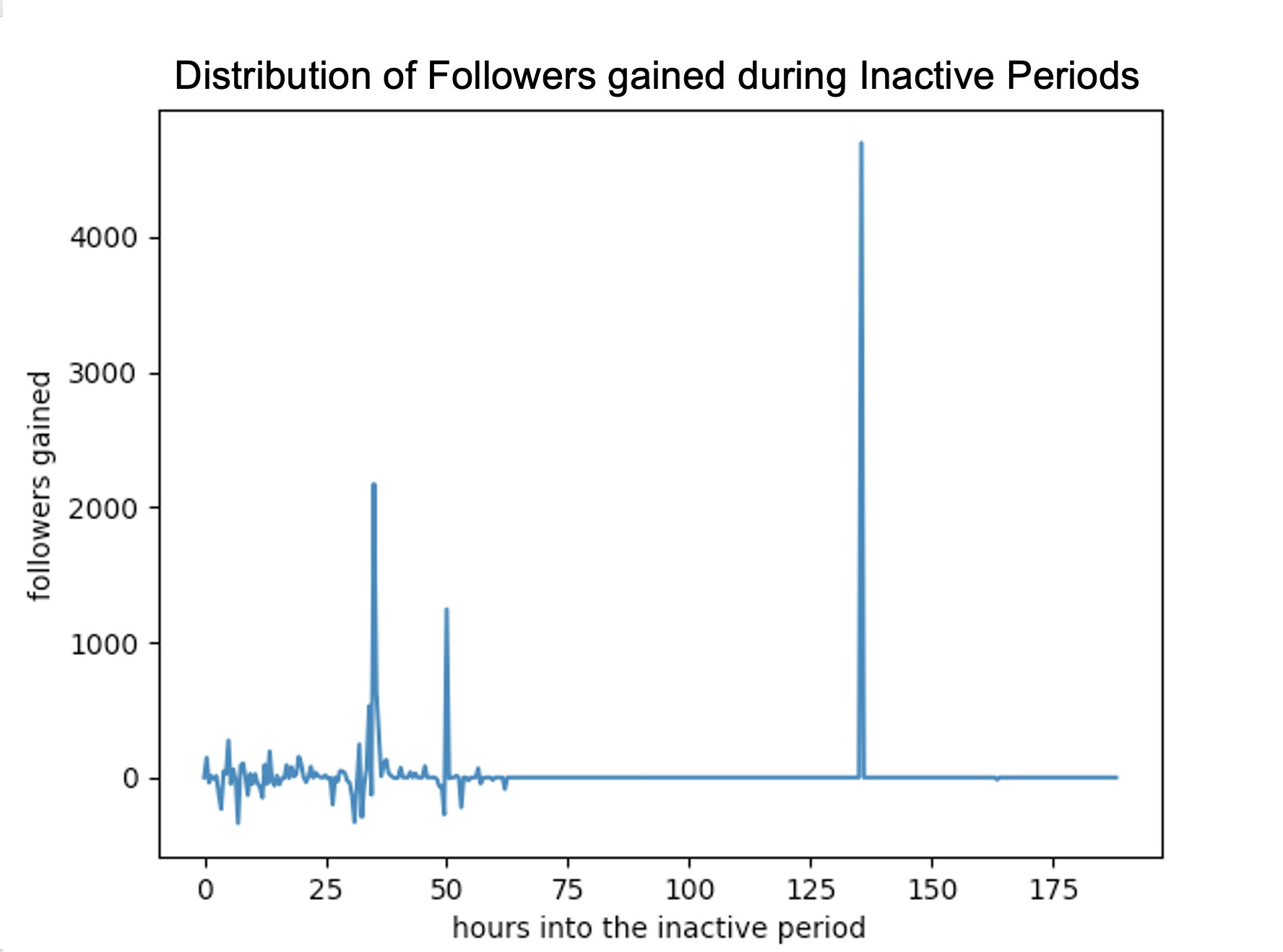}
\caption{Distribution of Followers Gained Inactive Period}
\end{minipage}
\end{figure}

To remedy the above problem, we employed a simple, naive algorithm to detect bot behaviors and delete users with such from our dataset. Our algorithm works by looking for sudden rise and drop in the number of followers between each 30 minutes timeframe. We calculated the absolute differences in followers between one time frame and the next. If the difference between timeframe $i$ and $i+1$ is at least 100 times larger than the difference between timeframe $i$ and $i-1$ and timeframe $i+1$ and $i+2$ for user A, we mark such user as abnormal and delete user A from the dataset. 

After running the algorithm on our set, we detected 226 users who have shown abnormal growth or decline during the inspected week, among which 110 streamers have more than 10,000 followers, and 67 of them have more than 100,000 followers. We provide a pseudocode of the algorithm below in Algorithm 1. 

\begin{algorithm}
    \DontPrintSemicolon 
    \KwIn{streamers dataset}
    \KwOut{streamers dataset w/o bot behavior}
    \For{userA in streamers dataset}{
    \For{timeframe[i] in userA}{
        \If{$|timeframe[i + 1] - timeframe[i]| > 100\cdot |timeframe[i] - timeframe[i - 1]|$ \textbf{and} $|timeframe[i + 1] - timeframe[i]| > 100\cdot |timeframe[i + 2] - timeframe[i + 1]|$}{mark userA as suspicious}
    }
        \If{userA is suspicious}{delete userA from the dataset}
    }
    \Return new dataset
    \caption{Bot Bahavior Detection}
\end{algorithm}

Even though heuristic algorithm is incomplete and inaccurate, it significantly helps us eschewing outliers and suspicious behaviors that tainted the results of the experiment. The result dataset after the application of bot detection shows major improvement on the distribution of followers growth during active and inactive period, which is shown in Figure 2.

The distribution of followers growth during active and inactive periods after applying our bot detection process is shown in Figure 6, 7. We observe that despite the occasional spikes in figure 6, there are not many fluctuations in the number of followers during inactive periods, which implies stability in the number of followers for most streamers when not streaming. A closer look at figure 7 reveals that most streamers experience the majority of increase in popularity during the first couple of hours of the streams; such increase starts to taper off when the streams go longer than 20 hours. 

\begin{figure}[h]
\begin{minipage}{16pc}
\includegraphics[width=14pc]{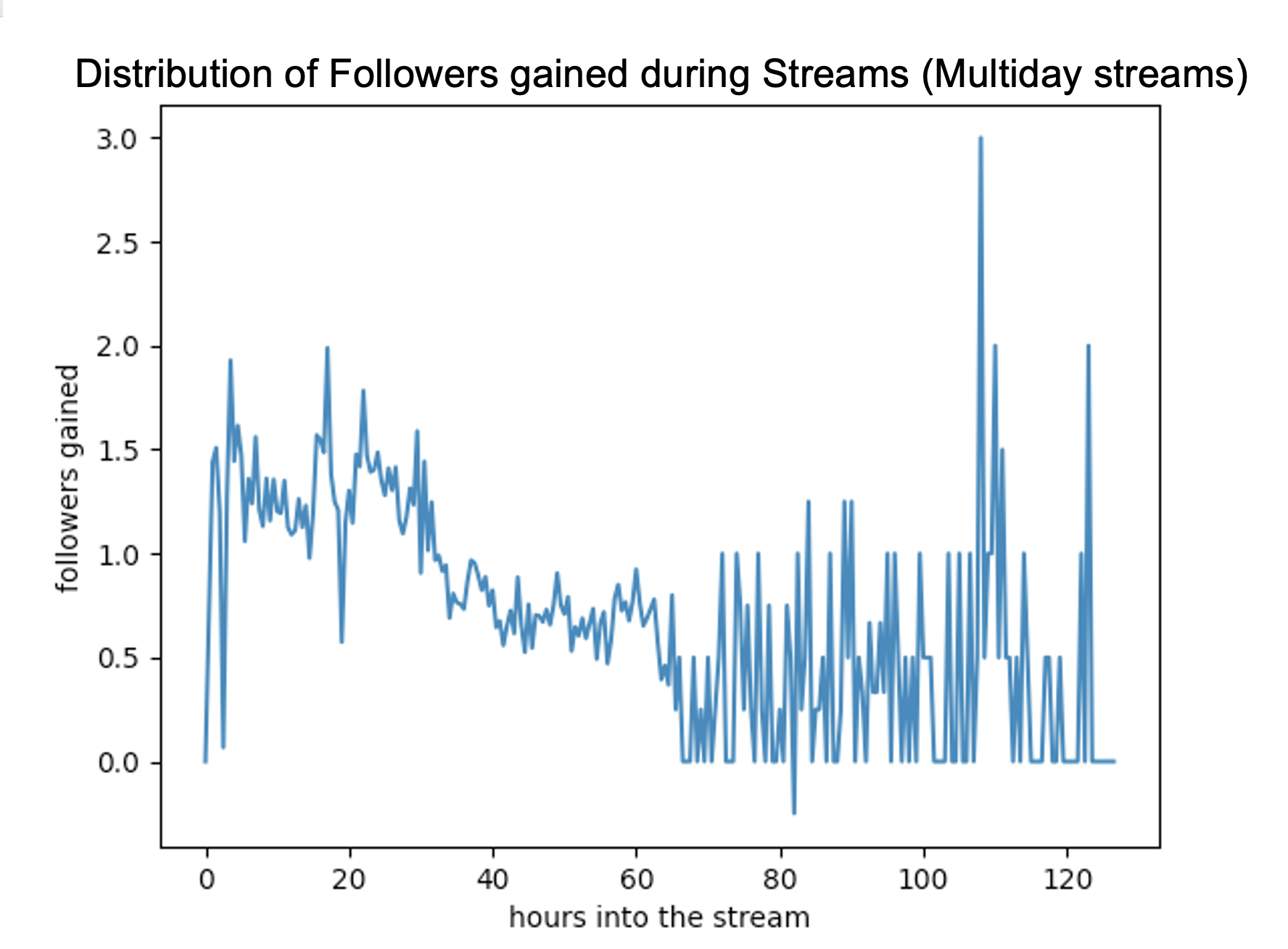}
\caption{Distribution of Followers Gained Active Period}
\end{minipage}\hspace{2pc}
\begin{minipage}{16pc}
\includegraphics[width=14pc]{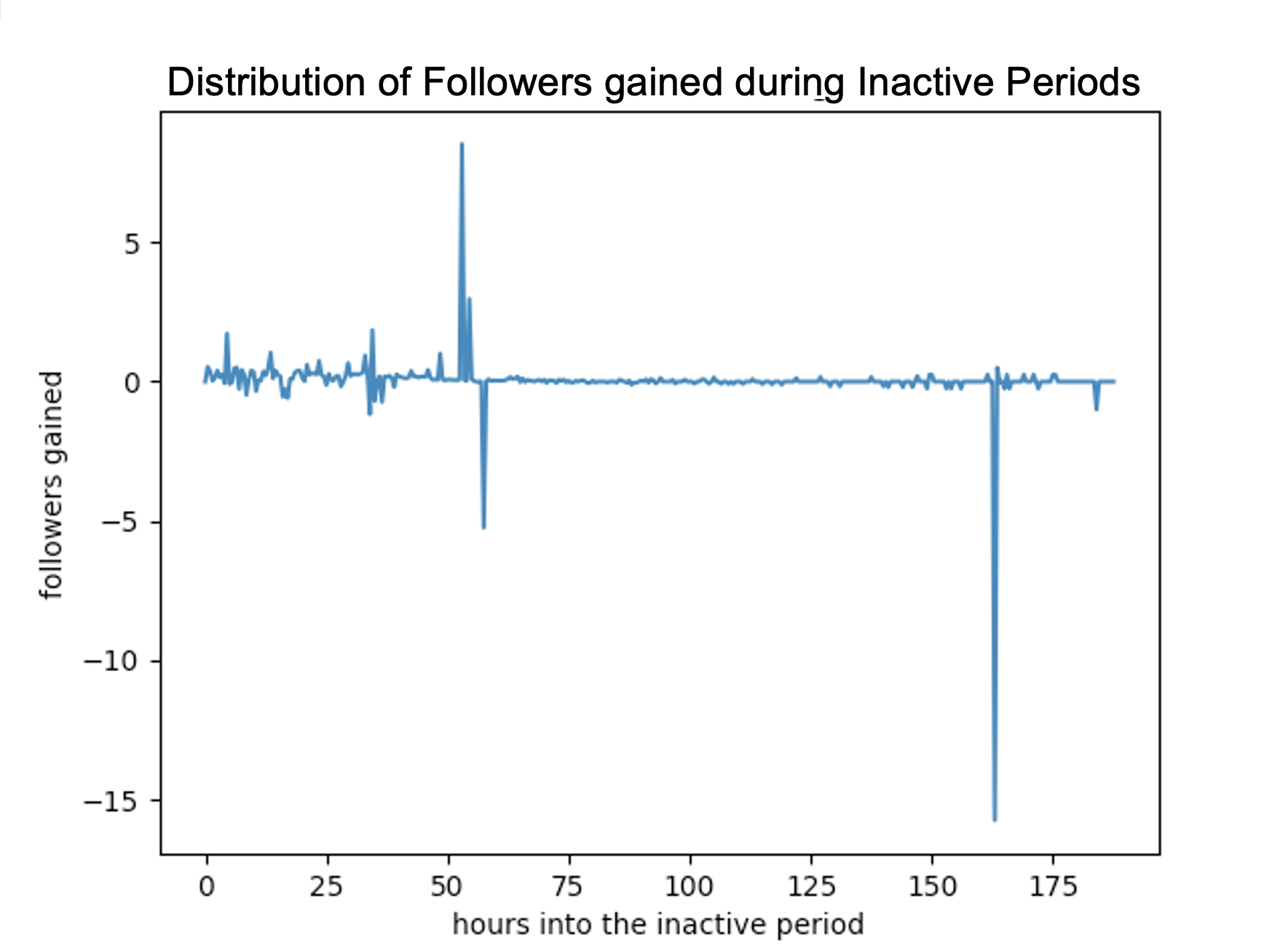}
\caption{Distribution of Followers Gained Inactive Period}
\end{minipage}
\end{figure}

\subsection{Categorization of Streamers}
Section 5.1 has shown that streamers' popularity can affect their ability to gain more followers/viewers. To identify other factors that can contribute to the success of a channel, we categorize the streamers in our dataset by their initial number of followers (the number of followers at the start of our observation period) into 4 categories (see Table 1).
\begin{table}
\caption{Categories of Users}
    \centering
    \begin{tabular}{c|c | c}
        \hline			
        \textbf{ Categories } & \textbf{ Initial number of followers } & \textbf{ Number of users }\\
        \hline
        small & 0-5000 & 1857 \\
        medium & 5000-10000 & 1177\\
        big & 10000-100000 & 3383\\
        mega & $>$ 100000 & 1062 \\
    \hline
    \end{tabular}
\end{table}

We mostly focus on small and medium streamers in our dataset since they are less susceptible to the rich gets richer effect discussed in Section 5.1. In the following Sections, we look at how streaming hours and the duration of streams can affect streamers' ability to gain more followers during active periods. 

\subsection{Duration of the Streams} Figure 8, 9 shows that the duration of a stream can affect the number of followers gained per streaming hour. For most users streaming less than 6 hours often brings the most optimal gain in followers per hour, while streams that last more than a day usually has less followers growth per hour. 

\begin{figure}[h]
\begin{minipage}{16pc}
\includegraphics[width=14pc, height=9pc]{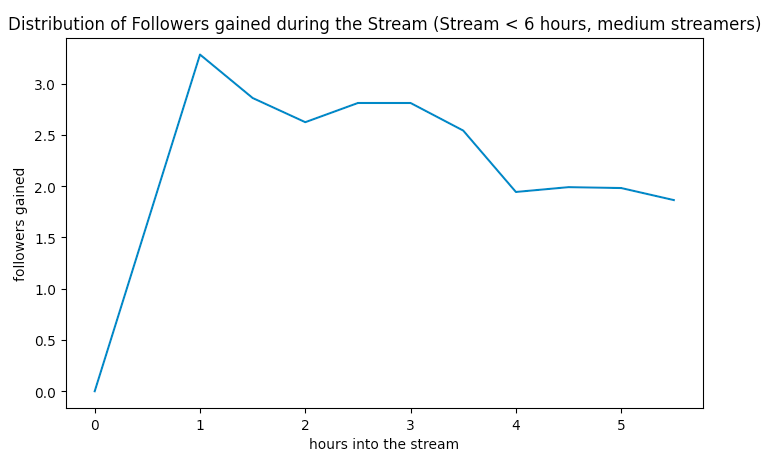}
\caption{Distribution of Followers Gained For Streams $<$ 6 hours}
\end{minipage}\hspace{2pc}
\begin{minipage}{16pc}
\includegraphics[width=14pc]{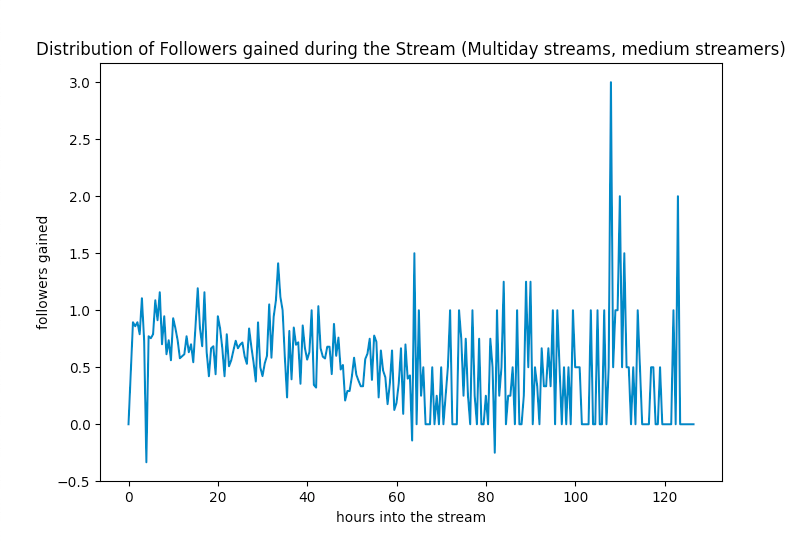}
\caption{Distribution of Followers Gained For Streams last $>$ 24 hours}
\end{minipage}
\end{figure}

\subsection{Streaming hours} Since streamers live and stream in different time zones, it is difficult to identify a specific optimal streaming hours for all users. Instead, we will identify users by their streaming languages and look for the best streaming time for each set. After analyzing the average followers gained for each 30 minutes timeframe, we noticed some trends. In particular, streams in some languages have several peak hours where streamers on average gain simnifically more followers, while streams in other languages only ave one peak hour. Figure 10 and 11 shows an example for each (English and Spanish). We will use UTC timezone for both graphs.

From Figure 10 and 11, we observe that streamers broadcasting in Spanish gain the most followers on average between 13:00 - 14:00 UTC time, while streamers broadcasting in English has several peak hours: 5:00 - 7:00, 16:00 - 17:00, and 20:00 - 21:00 UTC time. 

\begin{figure}[h]
\begin{minipage}{16pc}
\includegraphics[width=14pc]{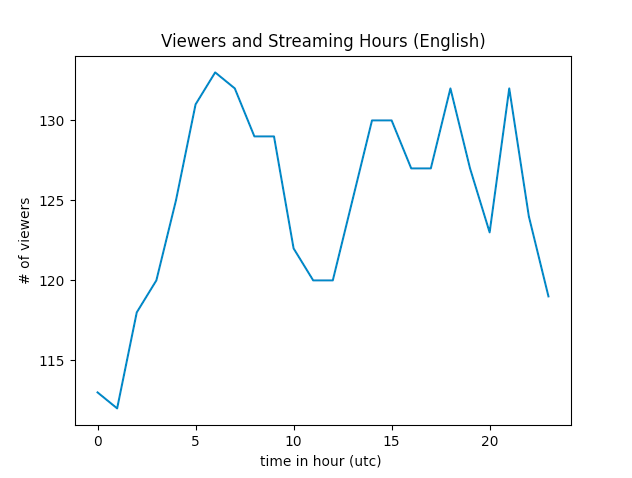}
\caption{Optimal Streaming Hours for English}
\end{minipage}\hspace{2pc}
\begin{minipage}{16pc}
\includegraphics[width=14pc]{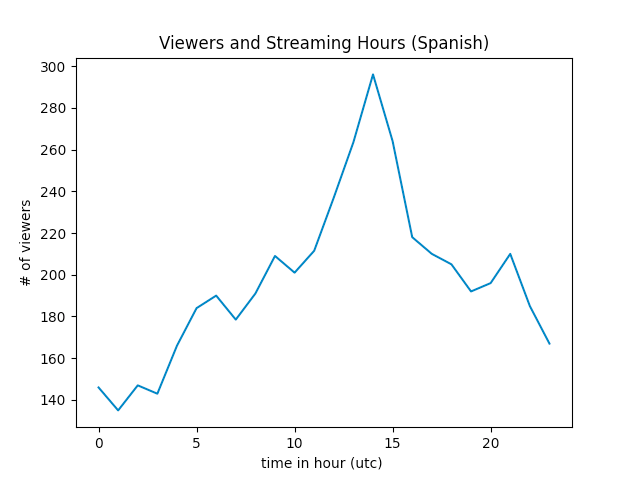}
\caption{Optimal Streaming Hours for Spanish}
\end{minipage}
\end{figure}

\subsection{Isolating Outliers}
In this Section, we employ a method to identify the best strategy for streamers to attract followers, outliers isolation. By identifying streamers who perform better than their peers, we hope to understand what they have in common in streaming characteristics.

First, we only consider streamers with initial number of followers between 0 and 20000. We then separate them into 20 "buckets" based on their initial popularity (the number of followers recorded at the start of our observation period): streamers with initially 0 - 1,000 followers, streamers with initially 1,000 - 2,000 followers, streamers with initially 2,000 - 3,000 followers, and so on, for up to streamers with initially 19,000 - 20,000 followers. For each bucket, we identify streamers who performed better than their peers by isolating the top 5$\%$ who gained the most subscribers. From the 5738 streamers being considered, we were able to identify 198 outliers. Next, we investigate what these outliers have in common and describe the result in the next few Sections. 

\subsection{Language Distribution} Figure 12 shows the pie chart of the distribution of streaming languages among the outliers. The most popular language used is English, followed by Spanish. This is to be expected since the most used languages among Twitch streamers are English and Spanish. We observe from the chart that all languages have at least one representative in the outliers. Thus, we conclude that languages used by streamers have little or no affect on channels' growth rate. 

\begin{figure}[h]
\begin{minipage}{16pc}
\includegraphics[width=14pc]{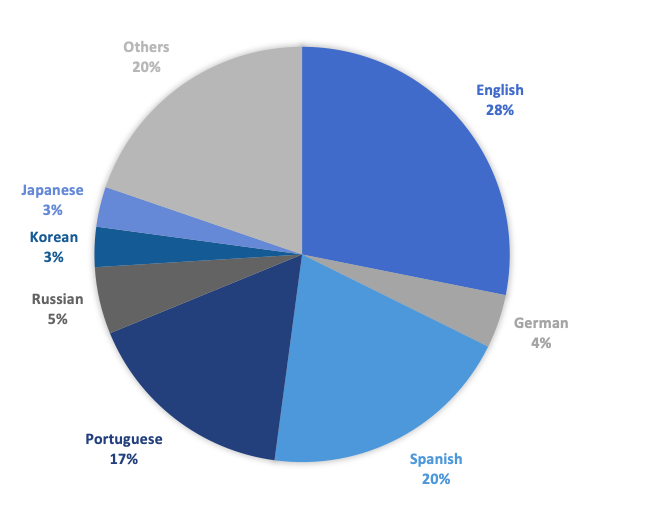}
\caption{Distribution of Languages}
\end{minipage}\hspace{2pc}
\begin{minipage}{16pc}
\includegraphics[width=14pc]{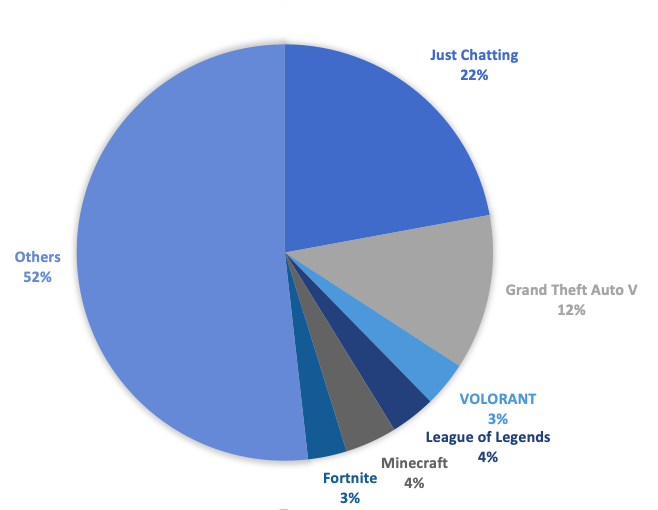}
\caption{Distribution of Content Tags}
\end{minipage}
\end{figure}

\subsection{Content Distribution} Figure 13 shows the pie chart for the types of gaming content streamed by outliers. Surprisingly, the most prominent type of content is Just Chatting, a non-gaming related tag, followed by the more well known gaming tags such as League of Legends, Grand Theft Auto V, and Minecraft. For a further investigation, we divided tags into two categories: gaming and non-gaming. Table 2 shows the percentages of gaming and non-gaming tags used by the outliers. We observe that 50$\%$ of the streams by the outliers have non-gaming tags, while only 29.22$\%$ of the streams in our entire dataset have non-gaming tags. This shows that successful streamers are more likely to stream a mix of non-gaming and gaming content.

\subsection{Length of the Stream}
Figure 14 and 15 show the distribution of the length of the streams for outliers and the entire dataset. We observe that most streams from the outliers generally stream much shorter streams in comparison to other streamers, with majority of the streams lasting less than 2 hours. For the entire population in our dataset, the majority of the streams last up to 6 hours, with many streams lasting for days. 

\begin{figure}[h]
\begin{minipage}{16pc}
\includegraphics[width=14pc]{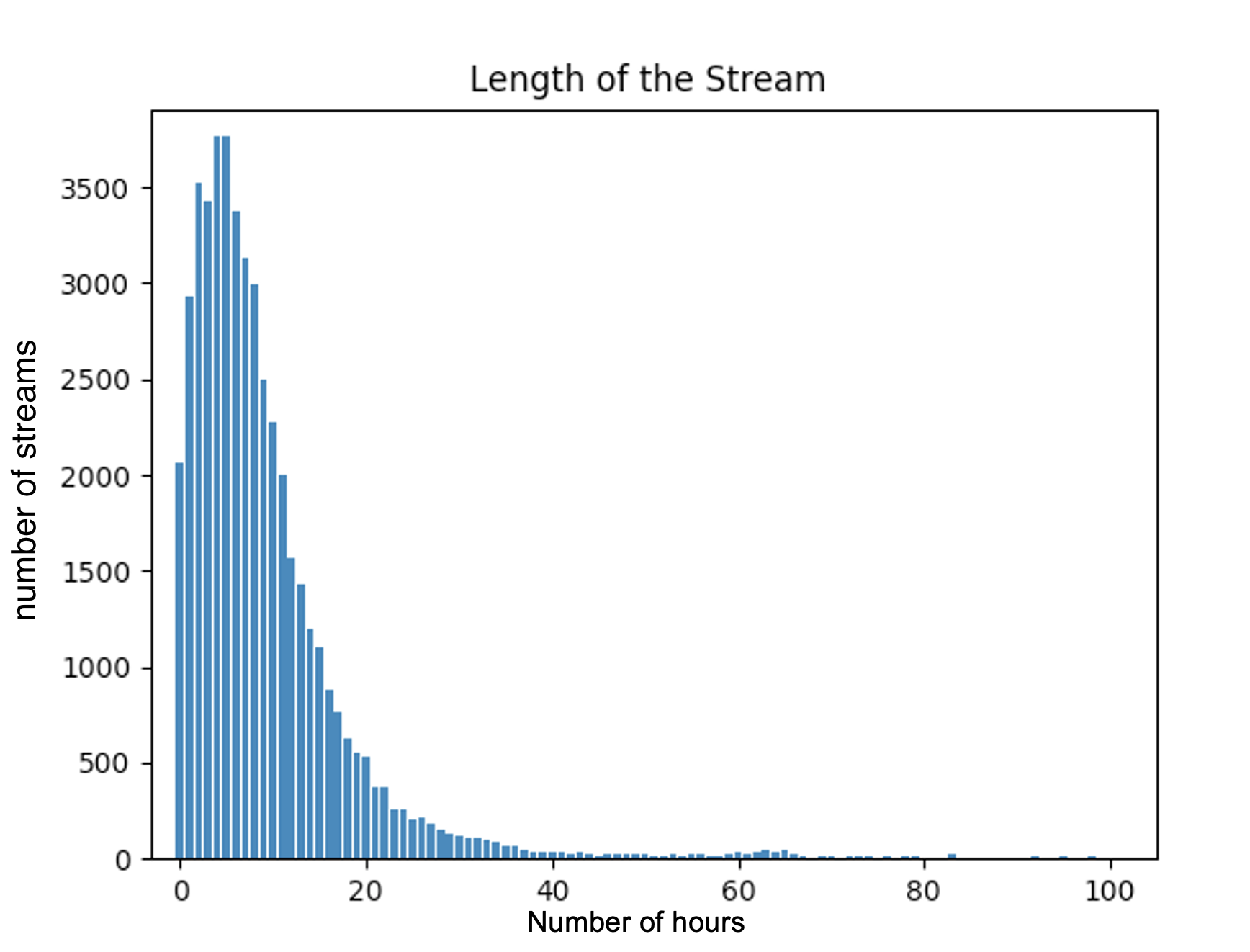}
\caption{Distribution of the Length of the Streams for the Entire Dataset}
\end{minipage}\hspace{2pc}
\begin{minipage}{16pc}
\includegraphics[width=14pc]{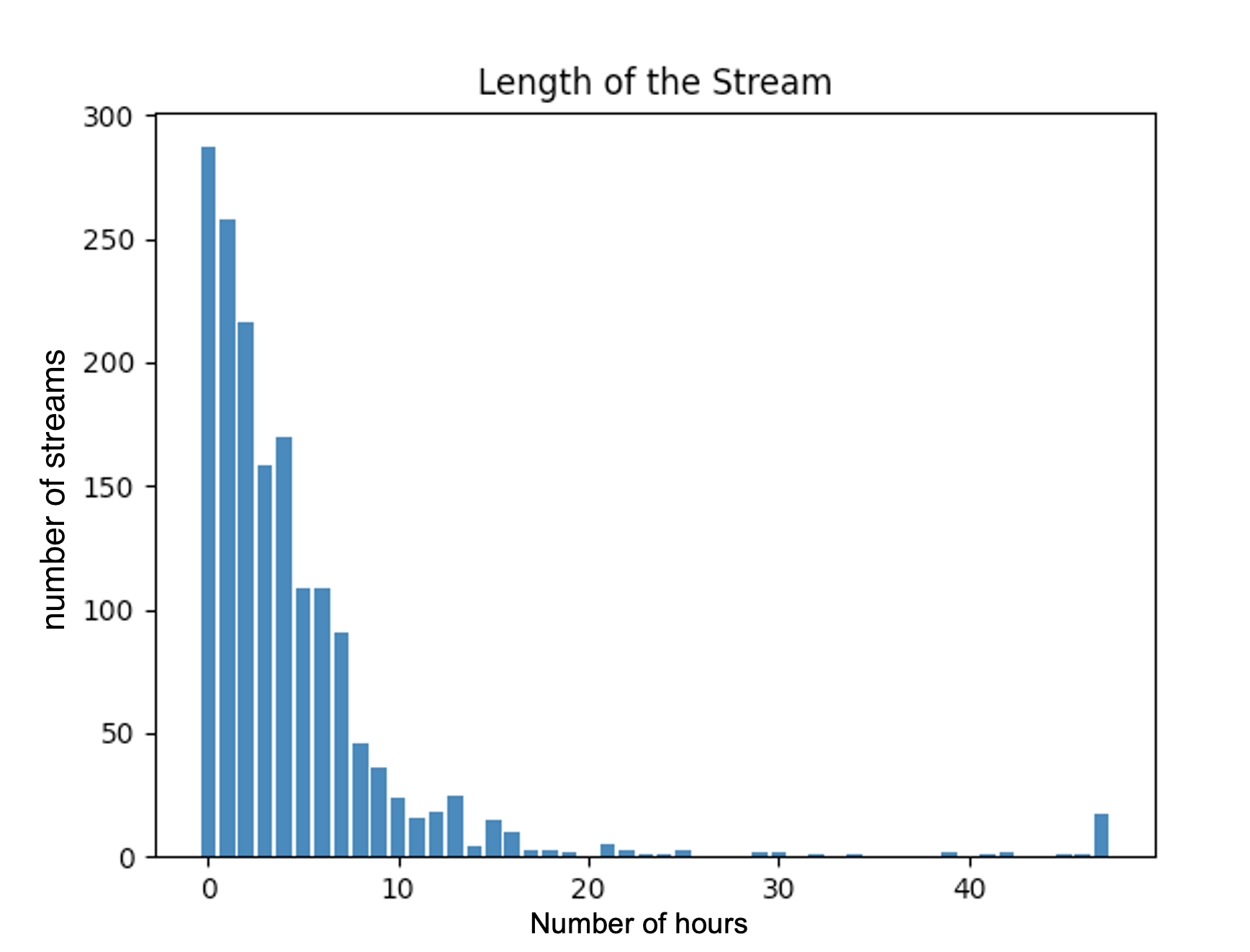}
\caption{Distribution of the Length of the Streams for the outliers}
\end{minipage}
\end{figure}

\subsection{Streaming Frequency}
Figure 16 and 17 show the distribution of streaming frequency for the outliers and for the entire dataset. We observe that most outliers stream more than 5 times per week, with many streaming between 5 - 8 times per week.

\begin{figure}[h]
\begin{minipage}{16pc}
\includegraphics[width=14pc]{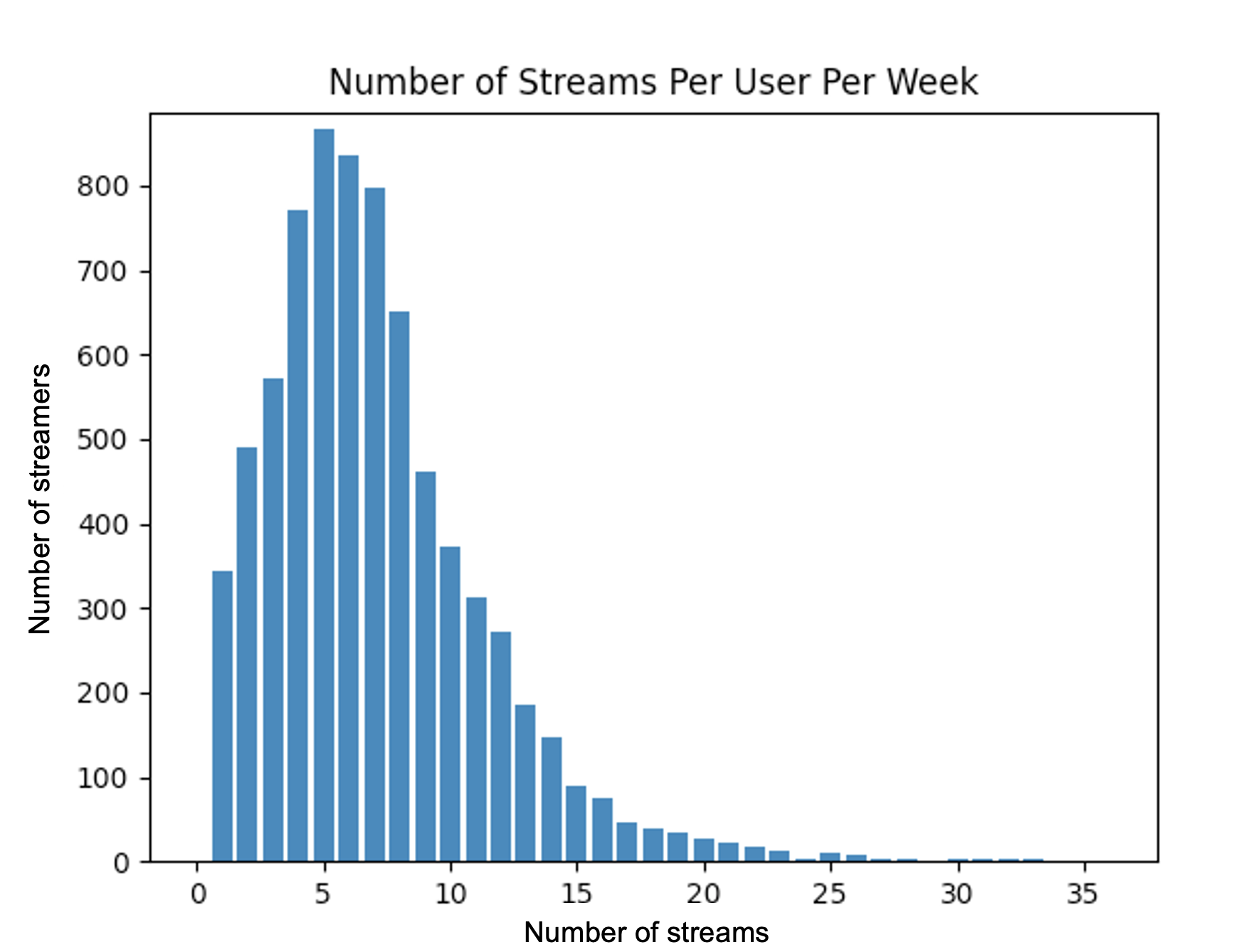}
\caption{Distribution of the $\#$ of Streams for the Entire Dataset}
\end{minipage}\hspace{2pc}
\begin{minipage}{16pc}
\includegraphics[width=14pc]{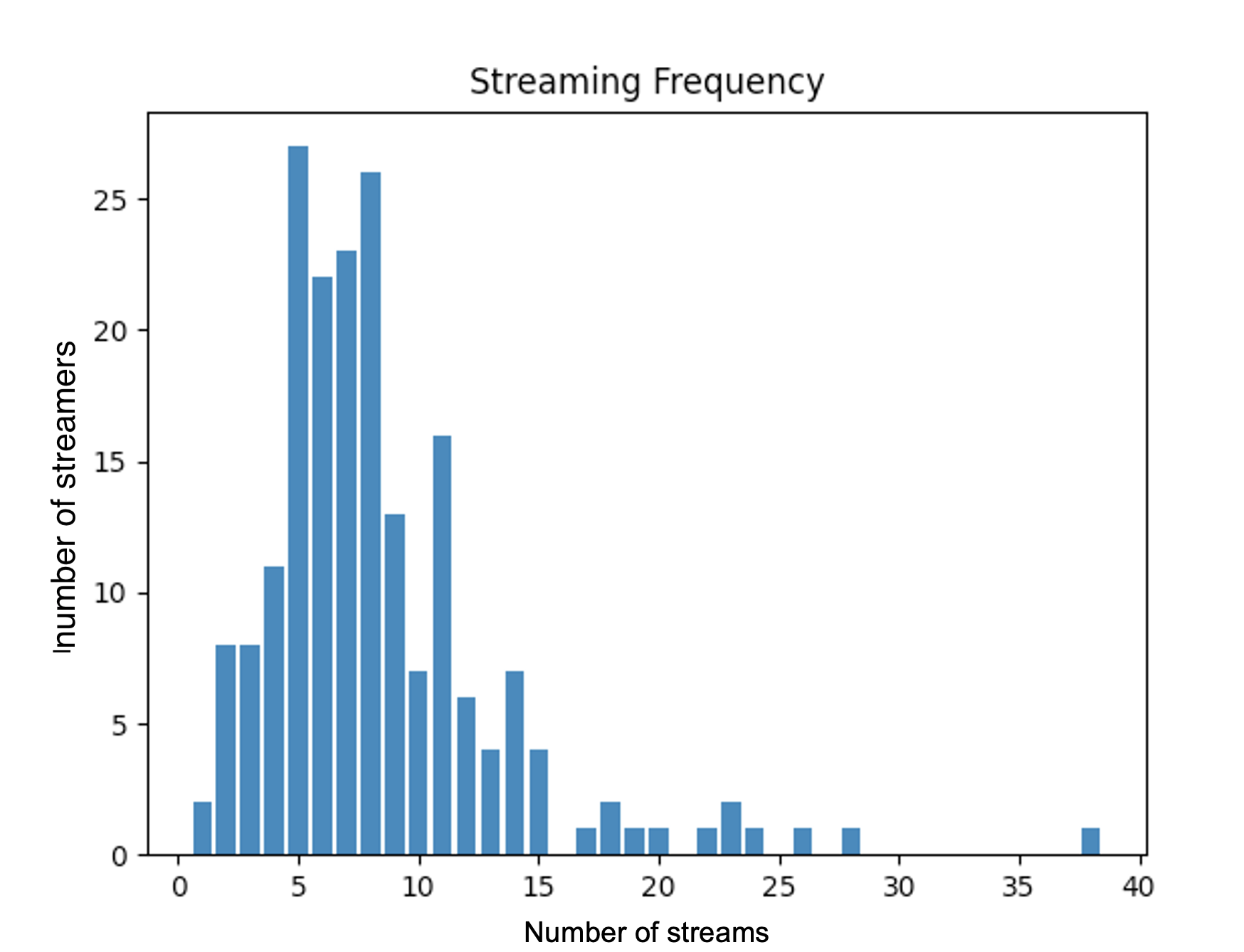}
\caption{Distribution of the $\#$ of Streams for the outliers}
\end{minipage}
\end{figure}

Overall, we observe from our results that a streaming strategy with multiple short streams and mixed content is the most effective. 

From the information above, we identify a common strategy of attracting more followers/viewers: 1) stream at least 5 times per week, 2) stream no more than 2 hours in each streaming, 3) gaming and non-gaming contents are mixed and evenly distributed as streaming contents, 4) stream more than 40 hours per week.
\begin{center}
\begin{table}[h]
\caption{Content Distribution}
\centering
\begin{tabular}{@{}l*{7}{l}}
\br
entire dataset($\%$)&outliers($\%$)\\
\mr
gaming & 70.78 & 50.00\\
non-gaming & 29.22 & 50.00\\
\br
\end{tabular}
\end{table}
\end{center}

Table 3 shows the statistics comparison between streamers apply that strategy and entire dataset. On average, streamers with that strategy gained triple as much followers as entire dataset.

\begin{center}
\begin{table}[h]
\caption{Comparison of followers gained}
\centering
\begin{tabular}{@{}l*{7}{l}}
\br
&Entire dataset&Streamers fit the strategy\\
\mr
Average Follower Gained & 101 & 342\\
Median Follower Gained & 29 & 67\\
Average Viewers per Hour & 101 & 101\\
\br
\end{tabular}
\end{table}
\end{center}

\subsection{Predicting the Growth of a Channel}
Our prediction task involves classifying users with initially 0 - 20k followers into classes based on their growth rate during the week. We aim to use users' quantitative streaming parameters to identify "good", "average", and "bad" channels (in reference to the channels' ability to attract followers). 

In our case, we chose to use a random forest classification model. Random forest classifier is a commonly used learning algorithm. Its mechanism involves the creation of multiple decision trees for training and the use of majority voting for the prediction process. For our study, we used the implementation provided in the Python Scikit Learn package. 

First, we classify channels based on their growth rates. We reused the bucket methods described in Section 5.4 and divided the channels into 20 buckets based on their initial number of followers. Then, if a channel looses followers during the observation process, we consider it a \textit{bad} channel. If a channel gained more followers than the average followers gained by channels in its bucket, we consider it a \textit{good} channel. Otherwise, the channel is marked as \textit{average}. Table 4 shows the average followers gained by channels in each bucket.

\begin{center}
\begin{table}[h]
\caption{Average followers gained for channels in each bucket}
\centering
\begin{tabular}{@{}l*{7}{l}}
\br
Initial followers&Average followers&Initial followers&Average followers\\
range&gained&range&gained\\
\mr
    0-1000 & 33.39 & 10000-11000 & 164.40\\
    1000-2000 & 61.74 & 11000-12000 & 205.43 \\
    2000-3000 & 77.60 & 12000-13000 & 207.54 \\
    3000-4000 & 76.16 & 13000-14000 & 133.64\\
    4000-5000 & 92.0 & 14000-15000 & 214.60\\
    5000-6000 & 117.29 & 15000-16000 & 309.73\\
    6000-7000 & 143.46 & 16000-17000 & 188.78\\
    7000-8000 & 139.39 & 17000-18000 & 127.06 \\
    8000-9000 & 159.08 & 18000-19000 & 359.18\\
    9000-10000 & 155.12 & 19000-20000 & 197.55\\
\br
\end{tabular}
\end{table}
\end{center}

Table 5 shows the size of each class. We can see that the classes are unbalanced. To improve on the accuracy of our model, we used oversampling to partially fix the imbalance of our dataset. We used Python Imblearn library's implementation of Synthetic Minority Oversampling Technique (SMOTE). We selected 80$\%$ of our data (4590 users) as the training set, and held 20$\%$ (1148 users) as the testing set. We apply oversampling to the training set to get a balanced dataset. Then, the oversampled training set is fed to our random forest model to learn. This model is then applied to the testing set to verify its accuracy. 
\begin{center}
\begin{table}[h]
\caption{Classes and class sizes}
\centering
\begin{tabular}{@{}l*{7}{l}}
\br
Class&Number of users\\
\mr
    good & 1339 \\
    average & 4213 \\
    bad & 186 \\
    total & 5738 \\
\br
\end{tabular}
\end{table}
\end{center}
We give the parameter fed to the model as well as the response variable the model returned. The learning parameters are: the initial number of followers and their bucket, the number of streams per week, the number of streams with less than 5 hours in length, the number of streams with 5 - 10 hours in length, the number of streams with more than 10 hours in length, gaming/non-gaming content (1 = yes, 0 = no), whether the streamers streamed popular games (1 = yes, 0 = no). They response variable is a prediction of whether the channel will perform good, bad or average (1 = good, 0 = average, -1 = bad). 

Lastly, we report on the accuracy of our trained model. We show the average accuracy, precision, recall and F1 measures of the model for 8 different maximum depths of the trees below. 

\begin{center}
\begin{table}[h]
\caption{Result of the model}
\centering
\begin{tabular}{@{}l*{7}{l}}
\br
Max depth&Accuracy($\%$)&Precision&Recall&f1\\
\mr
    2 & 40.06 & 0.61 & 0.40 & 0.43 \\
    4 & 54.61 & 0.63 & 0.54 & 0.58 \\
    6 & 66.63 & 0.65 & 0.66 & 0.65 \\
    8 & 71.34 & 0.66 & 0.71 & 0.67 \\
    10 & 71.51 & 0.66 & 0.71 & 0.67 \\
    12 & 70.73 & 0.65 & 0.70 & 0.66 \\
    14 & 68.55 & 0.63 & 0.68 & 0.64 \\
    16 & 66.81 & 0.62 & 0.66 & 0.64 \\
\br
\end{tabular}
\end{table}
\end{center}

From Table 6, we can see that the model with maximum trees' depth = 10 performs best with accuracy upto 71.55$\%$.
\section{Conclusions}
Over the past few years, streaming services have become an emerging source of media content. With the surge in popularity of e-sports and competitive gaming, Twitch has grappled the opportunity to become a major player in the gaming market. It hosts thousands of major gaming events and millions of independent streamers every year. With a customizable platform, it allows streamers and viewers to build their own communities and provide financial support for streamers. 

In this paper, we evaluate the different factors that could potentially contribute to streamers' success on Twitch. Our hope is to shed some light on the complex, intricate ecosystem of Twitch communities. 

Our data was collected using Twitch's API. We tracked the weekly activities of 20000 active Twitch users over the course of one month and gathered over 3M data points, with more than 55,00 streams and 50,000 videos, We inspected the features of streamers and their corresponding streams and identified which their importance. This includes the growth in followers, the number of views per video, the number of views per stream, and the content of the streams. 

Our model provides a tool for both Twitch communities and streamers. For Twitch communities, our model provides a guide to evaluate streamers and help identify future streaming celebrities. For streamers, we provide a self-evaluation method to help them with improving their streaming characteristics.  

\section*{References}
\nocite{*}
\bibliographystyle{unsrt}
\bibliography{isaic.bib}

\begin{thebibliography}{10}

\bibitem{borowy2013pioneering}
Michael Borowy et~al.
\newblock Pioneering esport: the experience economy and the marketing of early
  1980s arcade gaming contests.
\newblock {\em International Journal of Communication}, 7:21, 2013.

\bibitem{johnson2019impacts}
Mark~R Johnson and Jamie Woodcock.
\newblock The impacts of live streaming and twitch. tv on the video game
  industry.
\newblock {\em Media, Culture \& Society}, 41(5):670--688, 2019.

\bibitem{10.1145/2556288.2557048}
William~A. Hamilton, Oliver Garretson, and Andruid Kerne.
\newblock Streaming on twitch: Fostering participatory communities of play
  within live mixed media.
\newblock In {\em Proceedings of the SIGCHI Conference on Human Factors in
  Computing Systems}, CHI '14, page 1315–1324, New York, NY, USA, 2014.
  Association for Computing Machinery.

\bibitem{Churchill}
Benjamin~C.B. Churchill and Wen Xu.
\newblock The modem nation: A first study on twitch.tv social structure and
  player/game relationships.
\newblock In {\em 2016 IEEE International Conferences on Big Data and Cloud
  Computing (BDCloud), Social Computing and Networking (SocialCom), Sustainable
  Computing and Communications (SustainCom) (BDCloud-SocialCom-SustainCom)},
  pages 223--228, 2016.

\bibitem{greenwood2016social}
Shannon Greenwood, Andrew Perrin, and Maeve Duggan.
\newblock Social media update 2016.
\newblock {\em Pew Research Center}, 11(2):1--18, 2016.

\bibitem{dewing2010social}
Michael Dewing.
\newblock {\em Social media: An introduction}, volume~1.
\newblock Library of Parliament Ottawa, 2010.

\bibitem{qin2016political}
Bei Qin, David Str{\"o}mberg, and Yanhui Wu.
\newblock The political economy of social media in china.
\newblock Technical report, Working paper, 2016.

\bibitem{o2011impact}
Gwenn~Schurgin O'Keeffe, Kathleen Clarke-Pearson, et~al.
\newblock The impact of social media on children, adolescents, and families.
\newblock {\em Pediatrics}, 127(4):800--804, 2011.

\bibitem{siddiqui2016social}
Shabnoor Siddiqui, Tajinder Singh, et~al.
\newblock Social media its impact with positive and negative aspects.
\newblock {\em International Journal of Computer Applications Technology and
  Research}, 5(2):71--75, 2016.

\bibitem{edosomwan2011history}
Simeon Edosomwan, Sitalaskshmi~Kalangot Prakasan, Doriane Kouame, Jonelle
  Watson, and Tom Seymour.
\newblock The history of social media and its impact on business.
\newblock {\em Journal of Applied Management and entrepreneurship},
  16(3):79--91, 2011.

\bibitem{BBC}
Kim Gittleson.
\newblock Amazon buys video-game streaming site twitch.

\bibitem{livereport}
Ethan May.
\newblock Streamlabs \& stream hatchet q3 2020 live streaming industry report.

\bibitem{gamerdemographic}
Abraham Stein, Konstantin Mitgutsch, and Mia Consalvo.
\newblock Who are sports gamers? a large scale study of sports video game
  players.
\newblock {\em Convergence}, 19(3):345--363, 2013.

\bibitem{Tim2020}
Tim Wulf, Frank~M. Schneider, and Stefan Beckert.
\newblock Watching players: An exploration of media enjoyment on twitch.
\newblock {\em Games and Culture}, 15(3):328--346, 2020.

\bibitem{10.1145/3415165}
Jeff~T. Sheng and Sanjay~R. Kairam.
\newblock From virtual strangers to irl friends: Relationship development in
  livestreaming communities on twitch.
\newblock {\em Proc. ACM Hum.-Comput. Interact.}, 4(CSCW2), October 2020.

\bibitem{Zorah2018}
Zorah Hilvert-Bruce, James~T. Neill, Max Sjöblom, and Juho Hamari.
\newblock Social motivations of live-streaming viewer engagement on twitch.
\newblock {\em Computers in Human Behavior}, 84:58--67, 2018.

\bibitem{pires2015youtube}
Karine Pires and Gwendal Simon.
\newblock Youtube live and twitch: a tour of user-generated live streaming
  systems.
\newblock In {\em Proceedings of the 6th ACM multimedia systems conference},
  pages 225--230, 2015.

\bibitem{zhang2013understanding}
Boxun Zhang, Gunnar Kreitz, Marcus Isaksson, Javier Ubillos, Guido Urdaneta,
  Johan~A Pouwelse, and Dick Epema.
\newblock Understanding user behavior in spotify.
\newblock In {\em 2013 Proceedings IEEE INFOCOM}, pages 220--224. IEEE, 2013.

\bibitem{lee2016cannibalizing}
Minhyung Lee, HanByeol Choi, Daegon Cho, and Heeseok Lee.
\newblock Cannibalizing or complementing?. the impact of online streaming
  services on music record sales.
\newblock {\em Procedia Computer Science}, 91:662--671, 2016.

\bibitem{gandhi2021exploration}
Reesha Gandhi, Christine~L Cook, Nina LaMastra, Jirassaya Uttarapong, and
  Donghee~Yvette Wohn.
\newblock An exploration of mental health discussions in live streaming gaming
  communities.
\newblock {\em Frontiers in psychology}, 12:751, 2021.

\bibitem{todd2018gender}
Patricia~R Todd and Joanna Melancon.
\newblock Gender and live-streaming: source credibility and motivation.
\newblock {\em Journal of Research in Interactive Marketing}, 2018.

\bibitem{cai2021understanding}
Jie Cai, Cameron Guanlao, and Donghee~Yvette Wohn.
\newblock Understanding rules in live streaming micro communities on twitch.
\newblock In {\em Conference on Interactive Media Experiences (IMX’21)},
  2021.

\bibitem{cai2021moderation}
Jie Cai, Donghee~Yvette Wohn, and Mashael Almoqbel.
\newblock Moderation visibility: Mapping the strategies of volunteer moderators
  in live streaming micro communities.
\newblock 2021.

\bibitem{seering2018social}
Joseph Seering, Juan~Pablo Flores, Saiph Savage, and Jessica Hammer.
\newblock The social roles of bots: evaluating impact of bots on discussions in
  online communities.
\newblock {\em Proceedings of the ACM on Human-Computer Interaction},
  2(CSCW):1--29, 2018.

\bibitem{poyane2019toxic}
Roman Poyane.
\newblock Toxic communication on twitch. tv. effect of a streamer.
\newblock In {\em International Conference on Digital Transformation and Global
  Society}, pages 414--421. Springer, 2019.

\bibitem{kim2019subscriber}
Jina Kim, Kunwoo Bae, Eunil Park, and Angel~P. del Pobil.
\newblock Who will subscribe to my streaming channel? the case of twitch.
\newblock page 247–251. Association for Computing Machinery, 2019.

\bibitem{lee2019study}
Shao-Hung Lee.
\newblock A study of prediction of live-stream subscriptions: the case of
  twitch.
\newblock 2019.

\bibitem{han2016characteristics}
Sukhee Han.
\newblock Characteristics and comparison of popular channels on internet game
  broadcasting: Focus on twitch tv.
\newblock {\em The Journal of The Institute of Internet, Broadcasting and
  Communication}, 16(4):7--14, 2016.

\bibitem{Zhao}
Keran Zhao, Yuheng Hu, Kevin Hong, and J.~Westland.
\newblock Understanding characteristics of popular streamers on live streaming
  platforms: Evidence from twitch.tv.
\newblock {\em Journal of the Association for Information Systems}, 11 2020.

\bibitem{FigueiredoYouTubePatterns}
Flavio Figueiredo, Fabr\'{\i}cio Benevenuto, and Jussara~M. Almeida.
\newblock The tube over time: Characterizing popularity growth of youtube
  videos.
\newblock In {\em Proceedings of the Fourth ACM International Conference on Web
  Search and Data Mining}, WSDM '11, page 745–754, New York, NY, USA, 2011.
  Association for Computing Machinery.

\bibitem{PintoYouTubePrediction}
Henrique Pinto, Jussara~M. Almeida, and Marcos~A. Gon\c{c}alves.
\newblock Using early view patterns to predict the popularity of youtube
  videos.
\newblock In {\em Proceedings of the Sixth ACM International Conference on Web
  Search and Data Mining}, WSDM '13, page 365–374, New York, NY, USA, 2013.
  Association for Computing Machinery.

\bibitem{MaYoutubeLongterm}
Changsha Ma, Zhisheng Yan, and Chang~Wen Chen.
\newblock {\em LARM: A Lifetime Aware Regression Model for Predicting YouTube
  Video Popularity}, page 467–476.
\newblock Association for Computing Machinery, New York, NY, USA, 2017.

\bibitem{wu2015analyzing}
Bo~Wu and Haiying Shen.
\newblock Analyzing and predicting news popularity on twitter.
\newblock {\em International Journal of Information Management},
  35(6):702--711, 2015.

\bibitem{gao2019taxonomy}
Xiaofeng Gao, Zhenhao Cao, Sha Li, Bin Yao, Guihai Chen, and Shaojie Tang.
\newblock Taxonomy and evaluation for microblog popularity prediction.
\newblock {\em ACM Transactions on Knowledge Discovery from Data (TKDD)},
  13(2):1--40, 2019.

\bibitem{yu2020prediction}
Hai Yu, Ying Hu, and Peng Shi.
\newblock A prediction method of peak time popularity based on twitter
  hashtags.
\newblock {\em IEEE Access}, 8:61453--61461, 2020.

\bibitem{Soares2019Classify}
Carlos~Vicente Soares~Araujo, Marco~Antônio Pinheiro~de Cristo, and Rafael
  Giusti.
\newblock Predicting music popularity using music charts.
\newblock In {\em 2019 18th IEEE International Conference On Machine Learning
  And Applications (ICMLA)}, pages 859--864, 2019.

\bibitem{9434816}
Yutong Ge, Jiaqian Wu, and Yutong Sun.
\newblock Popularity prediction of music based on factor extraction and model
  blending.
\newblock In {\em 2020 2nd International Conference on Economic Management and
  Model Engineering (ICEMME)}, pages 1062--1065, 2020.

\bibitem{10.1145/3132847.3132997}
Changsha Ma, Zhisheng Yan, and Chang~Wen Chen.
\newblock {\em LARM: A Lifetime Aware Regression Model for Predicting YouTube
  Video Popularity}, page 467–476.
\newblock Association for Computing Machinery, New York, NY, USA, 2017.

\bibitem{Kong1}
Quyu Kong, Marian-Andrei Rizoiu, Siqi Wu, and Lexing Xie.
\newblock Will this video go viral: Explaining and predicting the popularity of
  youtube videos.
\newblock In {\em Companion Proceedings of the The Web Conference 2018}, WWW
  '18, page 175–178, Republic and Canton of Geneva, CHE, 2018. International
  World Wide Web Conferences Steering Committee.

\bibitem{jagongo2013social}
Ambrose Jagongo and Catherine Kinyua.
\newblock The social media and entrepreneurship growth.
\newblock {\em International journal of humanities and social science},
  3(10):213--227, 2013.

\bibitem{hajli2014study}
M~Nick Hajli.
\newblock A study of the impact of social media on consumers.
\newblock {\em International Journal of Market Research}, 56(3):387--404, 2014.

\bibitem{ewalt2013espn}
David~M Ewalt.
\newblock The espn of videogames, 2013.

\bibitem{hoiles2017engagement}
William Hoiles, Anup Aprem, and Vikram Krishnamurthy.
\newblock Engagement and popularity dynamics of youtube videos and sensitivity
  to meta-data.
\newblock {\em IEEE Transactions on Knowledge and Data Engineering},
  29(7):1426--1437, 2017.

\bibitem{lewinski2015don}
Peter Lewinski.
\newblock Don’t look blank, happy, or sad: Patterns of facial expressions of
  speakers in banks’ youtube videos predict video’s popularity over time.
\newblock {\em Journal of Neuroscience, Psychology, and Economics}, 8(4):241,
  2015.

\bibitem{welbourne2016science}
Dustin~J Welbourne and Will~J Grant.
\newblock Science communication on youtube: Factors that affect channel and
  video popularity.
\newblock {\em Public understanding of science}, 25(6):706--718, 2016.

\bibitem{Wong1}
Ekapol Wongsuparatkul and Sukree Sinthupinyo.
\newblock View count of online videos prediction using clustering view count
  patterns with multivariate linear model.
\newblock In {\em Proceedings of the 8th International Conference on Computer
  and Communications Management}, ICCCM'20, page 123–129, New York, NY, USA,
  2020. Association for Computing Machinery.

\bibitem{bartl2018youtube}
Mathias B{\"a}rtl.
\newblock Youtube channels, uploads and views: A statistical analysis of the
  past 10 years.
\newblock {\em Convergence}, 24(1):16--32, 2018.

\bibitem{de2017predicting}
Shaunak De, Abhishek Maity, Vritti Goel, Sanjay Shitole, and Avik Bhattacharya.
\newblock Predicting the popularity of instagram posts for a lifestyle magazine
  using deep learning.
\newblock In {\em 2017 2nd International Conference on Communication Systems,
  Computing and IT Applications (CSCITA)}, pages 174--177. IEEE, 2017.

\bibitem{park2016style}
Jaehyuk Park, Giovanni~Luca Ciampaglia, and Emilio Ferrara.
\newblock Style in the age of instagram: predicting success within the fashion
  industry using social media.
\newblock In {\em Proceedings of the 19th ACM Conference on computer-supported
  cooperative work \& social computing}, pages 64--73, 2016.

\bibitem{bakhshi2014faces}
Saeideh Bakhshi, David~A Shamma, and Eric Gilbert.
\newblock Faces engage us: Photos with faces attract more likes and comments on
  instagram.
\newblock In {\em Proceedings of the SIGCHI conference on human factors in
  computing systems}, pages 965--974, 2014.

\bibitem{han2018photos}
Kyungsik Han, Yonggeol Jo, Youngseung Jeon, Bogoan Kim, Junho Song, and
  Sang-Wook Kim.
\newblock Photos don't have me, but how do you know me? analyzing and
  predicting users on instagram.
\newblock In {\em Adjunct publication of the 26th conference on user modeling,
  adaptation and personalization}, pages 251--256, 2018.

\bibitem{carta2020popularity}
Salvatore Carta, Alessandro~Sebastian Podda, Diego~Reforgiato Recupero, Roberto
  Saia, and Giovanni Usai.
\newblock Popularity prediction of instagram posts.
\newblock {\em Information}, 11(9):453, 2020.

\bibitem{zohourian2018popularity}
Alireza Zohourian, Hedieh Sajedi, and Arefeh Yavary.
\newblock Popularity prediction of images and videos on instagram.
\newblock In {\em 2018 4th International Conference on Web Research (ICWR)},
  pages 111--117. IEEE, 2018.

\bibitem{yazdani2015predicting}
Mehrdad Yazdani and Lev Manovich.
\newblock Predicting social trends from non-photographic images on twitter.
\newblock In {\em 2015 IEEE international conference on big data (big data)},
  pages 1653--1660. IEEE, 2015.

\bibitem{jaidka2019predicting}
Kokil Jaidka, Saifuddin Ahmed, Marko Skoric, and Martin Hilbert.
\newblock Predicting elections from social media: a three-country, three-method
  comparative study.
\newblock {\em Asian Journal of Communication}, 29(3):252--273, 2019.

\bibitem{zhang2011predicting}
Xue Zhang, Hauke Fuehres, and Peter~A Gloor.
\newblock Predicting stock market indicators through twitter “i hope it is
  not as bad as i fear”.
\newblock {\em Procedia-Social and Behavioral Sciences}, 26:55--62, 2011.

\bibitem{chung2011predicting}
Sang Chung and Sandy Liu.
\newblock Predicting stock market fluctuations from twitter.
\newblock {\em Berkeley, California}, 2011.

\bibitem{golbeck2011predicting}
Jennifer Golbeck, Cristina Robles, Michon Edmondson, and Karen Turner.
\newblock Predicting personality from twitter.
\newblock In {\em 2011 IEEE third international conference on privacy,
  security, risk and trust and 2011 IEEE third international conference on
  social computing}, pages 149--156. IEEE, 2011.

\bibitem{asur2010predicting}
Sitaram Asur and Bernardo~A Huberman.
\newblock Predicting the future with social media.
\newblock In {\em 2010 IEEE/WIC/ACM international conference on web
  intelligence and intelligent agent technology}, volume~1, pages 492--499.
  IEEE, 2010.

\bibitem{ruwantha2020lstm}
WMDR Ruwantha and BTGS Kumara.
\newblock Lstm based approach for classifying twitter posts for movie success
  prediction.
\newblock In {\em 2020 International Conference on Decision Aid Sciences and
  Application (DASA)}, pages 1160--1165. IEEE, 2020.

\bibitem{farrington2015analysis}
Daniel Farrington and N~Muesch.
\newblock Analysis of the charisteristics and content of twitch live-streaming.
\newblock {\em Interactive Qualifying Project IQP-MLC-TT14}, 2015.

\bibitem{LatermanCampusview}
Michel Laterman, Martin Arlitt, and Carey Williamson.
\newblock A campus-level view of netflix and twitch: Characterization and
  performance implications.
\newblock In {\em 2017 International Symposium on Performance Evaluation of
  Computer and Telecommunication Systems (SPECTS)}, pages 1--8, 2017.

\bibitem{dux2018social}
James Dux and Janghyun Kim.
\newblock Social live-streaming: Twitch. tv and uses and gratification theory
  social network analysis.
\newblock {\em Computer Science \& Information Technology}, 47, 2018.

\end{thebibliography}

\end{document}